\documentclass[pra,final,twocolumn,amsmath,eqsecnum,amssymb,superscriptaddress,amsfonts]{revtex4-1}
\usepackage{graphicx,graphics}
\usepackage{color}

\usepackage{epsfig}
\usepackage{amsmath}
\usepackage{amssymb}
\usepackage{bm}

\usepackage{textcomp}

\begin{document}

\title{$d-p$ model and spin-orbital order in the vanadium perovskites}

\author{     Krzysztof Ro\'sciszewski}
\affiliation{\mbox{Marian Smoluchowski Institute of Physics, Jagiellonian University,
             Prof. S. \L{}ojasiewicza 11, PL-30348 Krak\'ow, Poland}}

\author{     Andrzej M. Ole\'s  }
\affiliation{\mbox{Marian Smoluchowski Institute of Physics, Jagiellonian University,
             Prof. S. \L{}ojasiewicza 11, PL-30348 Krak\'ow, Poland}}
\affiliation{Max Planck Institute for Solid State Research,
             Heisenbergstrasse 1, D-70569 Stuttgart, Germany }

\date{\today}

\begin{abstract}
Using the multi-band $d-p$ model and unrestricted Hartree-Fock
approximation we investigate the electronic structure and spin-orbital
order in three-dimensional VO$_3$ lattice. The main aim of this
investigation is testing if simple $d-p$ model, with partly filled
$3d$ orbitals (at vanadium ions) and $2p$ orbitals (at oxygen ions),
is capable of reproducing correctly nontrivial coexisting spin-orbital
order observed in the vanadium perovskites. We point out that the
multi-band $d-p$ model has to include partly filled $e_g$ orbitals at
vanadium ions. The results suggest weak self-doping as an important
correction beyond the ionic model and reproduce the possible ground
states with broken spin-orbital symmetry on vanadium ions:
either $C$-type alternating orbital order accompanied by $G$-type
antiferromagnetic spin order,
or $G$-type alternating orbital order accompanied by $C$-type
antiferromagnetic spin order.
Both states are experimentally observed and compete with each other in
YVO$_3$ while only the latter was observed in LaVO$_3$. Orbital order
is induced and stabilized by particular patterns of oxygen distortions
arising from the Jahn-Teller effect. In contrast to time-consuming
\textit{ab-initio} calculations, the computations using $d-p$ model
are very quick and should be regarded as very useful in solid state
physics, provided the parameters are selected carefully.
\end{abstract}


\maketitle

\section{Introduction}
\label{intro}

The spin and orbital ordering found in three dimensional (3D) vanadium
perovskites is an old but still very interesting problem with many
challenges. It was discussed in numerous experimental and theoretical
papers, considering undoped \cite{Miz99,Kaw94,Ngu95,Saw96,Nak99,Ren00,Bla01,Kha04,Kha01,Hor03,Jo03,
Ren03,Fan04,Ulr03,Ots06,Ree06,Sol06,Ray07,Hor08,Mos09,Mos10,Fuj10,Kum17,
Kim18} and doped by charged defects \cite{Fuj05,Ave13,Ave15} vanadium
perovskites.
On the theoretical side, the first insightful explanation of the
alternating orbital (AO) order was given by Mizokawa, Khomskii, and
Sawatzky in 1999 \cite{Miz99}. They studied the competition between two
types of spin-orbital order in vanadates within the so-called lattice
model. It was claimed that Jahn-Teller (JT) distortions of the lattice
\cite{Kan60} (see Fig.~1) are primarily responsible for the onset of
this order. Sizable tilting of the apical axes of octahedra (out of an
ideal cubic structure) was assumed to be the main driving factor which
distinguishes between low temperature and high temperature
order in LaVO$_3$ or YVO$_3$ \cite{Miz99}.

\begin{figure}[t!]
\vskip -1.5cm
\begin{center}
\includegraphics[width=12.4cm]{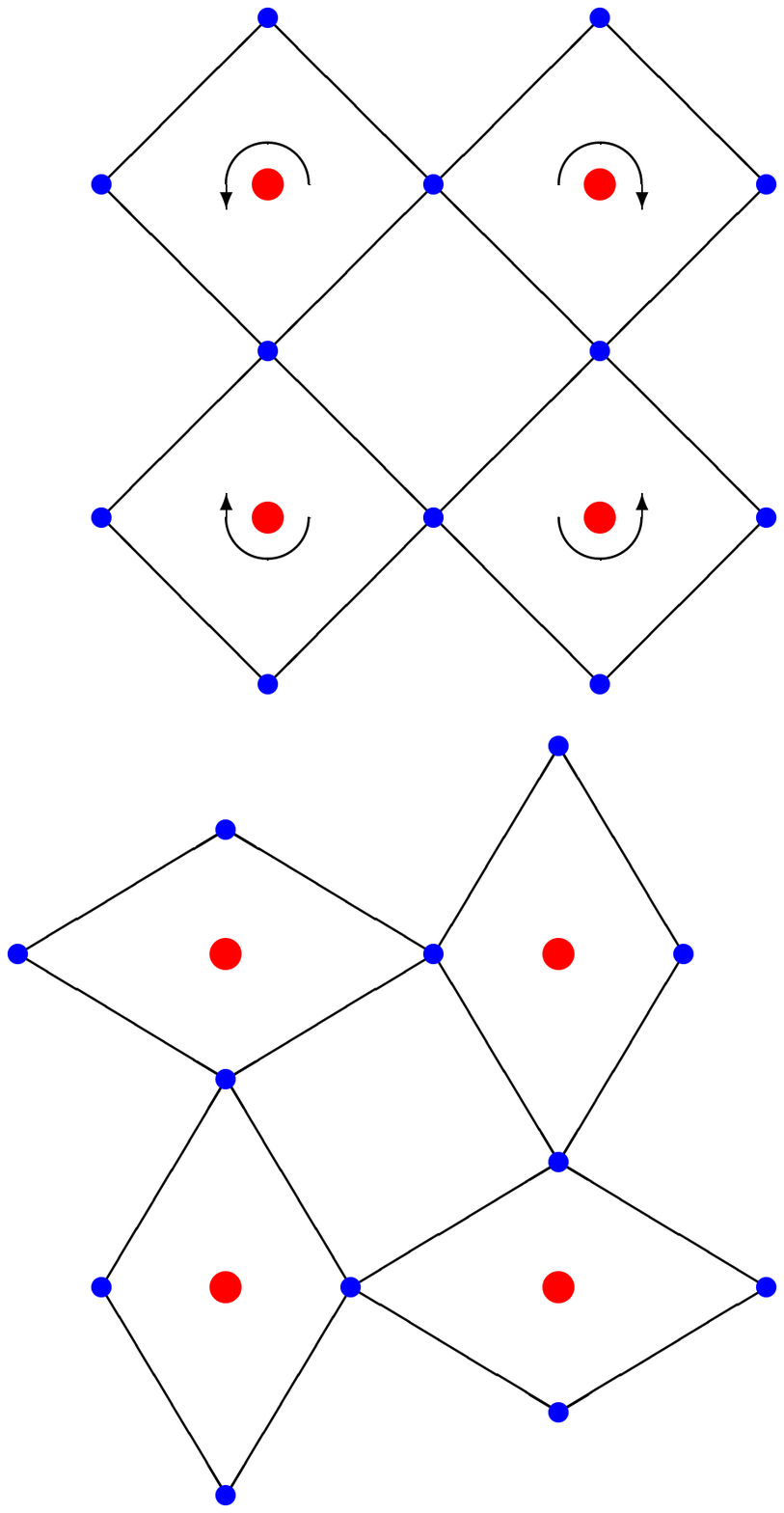}
\end{center}
\vspace{-4.4cm}
\caption{Schematic view of {\it cooperative and static} $Q_4$ JT
distortions involving rotations of octahedra  groups (upper panel)
and $Q_2$ distortions (lower panel). For description and classification
of different JT modes see Ref. \cite{Kan60}. Red/blue dots denote
positions of vanadium/oxygen ions in the $ab$ plane.
}
\label{fig1}
\end{figure}

Easy-to-grasp presentation of the spin and orbital order in the ground
state as perceived today by experimentalists was presented by Blake
\textit{et al.} \cite{Bla01}. The phase diagram of the vanadium
perovskites $R$VO$_3$ \cite{Fuj10} shows several spin- and/or orbital
ordered phases. In the regime of compounds with low values of ionic
radii $r_R$ of rare earth ions $R$ as in YVO$_3$, two antiferromagnetic
(AF) phases with complementary spin-orbital order appear:
(i)~$G$-type AF ($G$-AF) order accompanied by $C$-type
alternating orbital order ($C$-AO)
with staggered orbitals in $ab$ planes and repeated orbitals
along the $c$ axis (below the magnetic transition at $T_{\rm N2}=77$ K)
and
(ii) $C$-type AF ($C$-AF) order accompanied by $G$-type AO ($G$-AO)
order for $T_{\rm N2}<T<T_{\rm N1}$, where $T_{\rm N1}=116$ K is the
high-temperature magnetic transition \cite{Fuj10}.

\begin{figure}[t!]
\vskip -1.5cm
\begin{center}
\includegraphics[width=12.4cm]{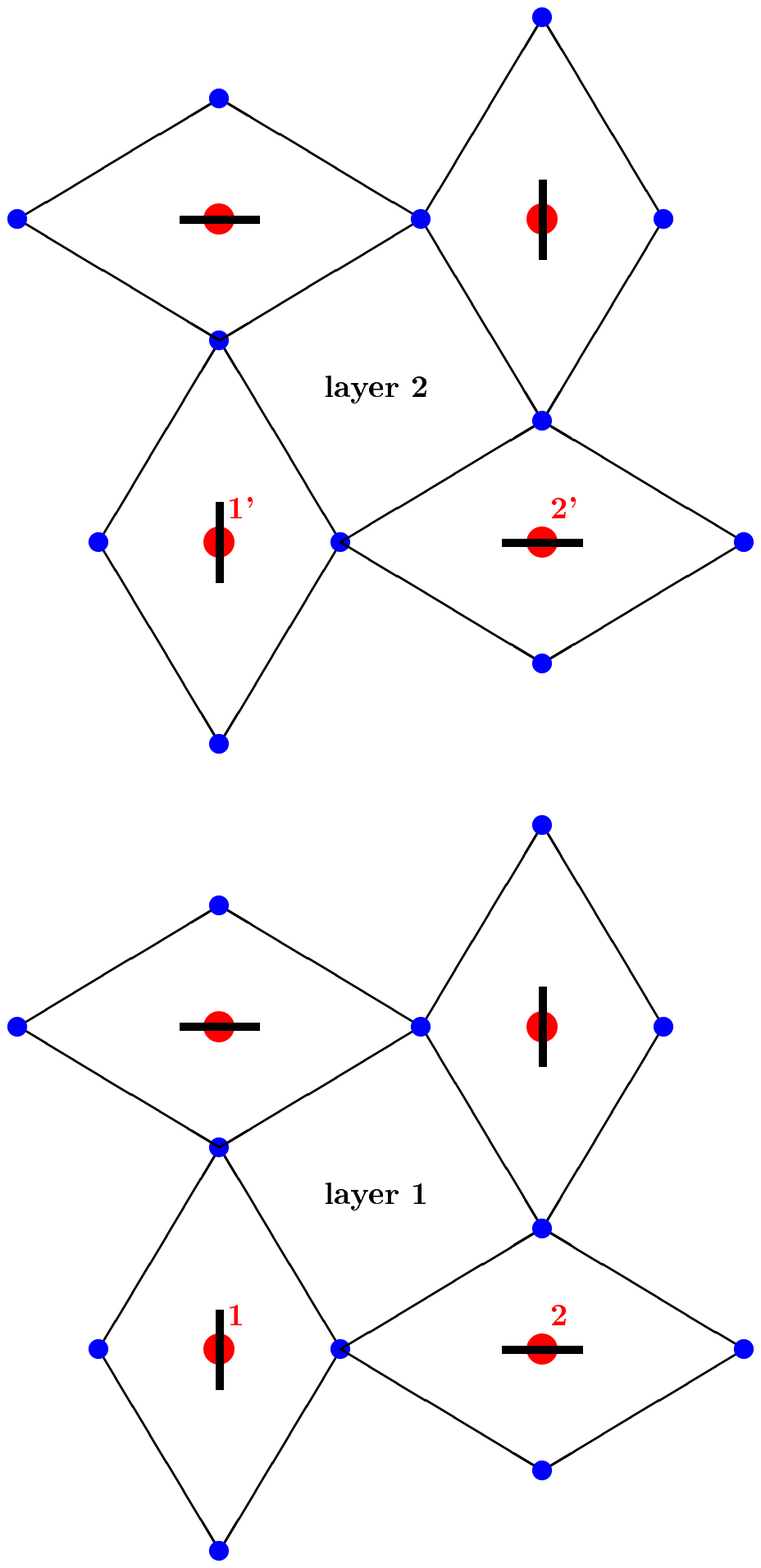}
\end{center}
\vspace{-4.0cm}
\caption{%
Schematic view of JT distortions used for Hartree-Fock computations in
the low-temperature phase of YVO$_3$. The long bars denote preferred
$yz$ or $zx$ orbitals --- their cooperative arrangement forms $C$-AO
order. Spins are not shown. The numbers shown close to vanadium positions
identify the ions (see the corresponding entries in Table II).
Horizontal and vertical directions on the figure correspond to $x$ and
$y$ axes, respectively; note that the $x,y$ axes are at 45 $\deg$ angle
to the crystallographic $a,b$ axes, i.e., our $x$ direction  corresponds
to crystallographic  (1,1,0)  direction.
The orbital order is repeated in consecutive layers when moving up
along the $z$ axis (this coincides with crystallographic $c$ axis).
}
\label{fig2}
\end{figure}

It is well understood now that at zero temperature, i.e., when YVO$_3$
is orthorhombic, the $zx$ and $yz$ orbitals on vanadium ions alternate
between two sublattices forming orbital $C$-AO long range order and
this order resembles AF spin order in a single $ab$ plane, see Fig. 2,
while along the $c$ axis this order is repeated, i.e., there is an
analogy to ordinary spin ferromagnetic (FM) order \cite{Bla01}. At the
same time the spins are arranged according to ordinary 3D N\'eel state
($G$-AF spin order). At intermediate temperatures $T>77$ K
(when YVO$_3$ is monoclinic) this order is reversed: the $G$-AO order
is accompanied by $C$-AF spin order, see Fig. 3.
The magnetic transition at $T=77$ K is triggered by the dimerization
in spin-orbital chains which requires spin fluctuations at finite
temperature \cite{Sir08}. Altogether this transition takes place
between two types of spin-orbital order along the $c$ axis which
follow the complementarity predicted by the Goodenough-Kanamori rules
\cite{Goode}.

\begin{figure}[t!]
\vskip -1.5cm
\begin{center}
\includegraphics[width=12.4cm]{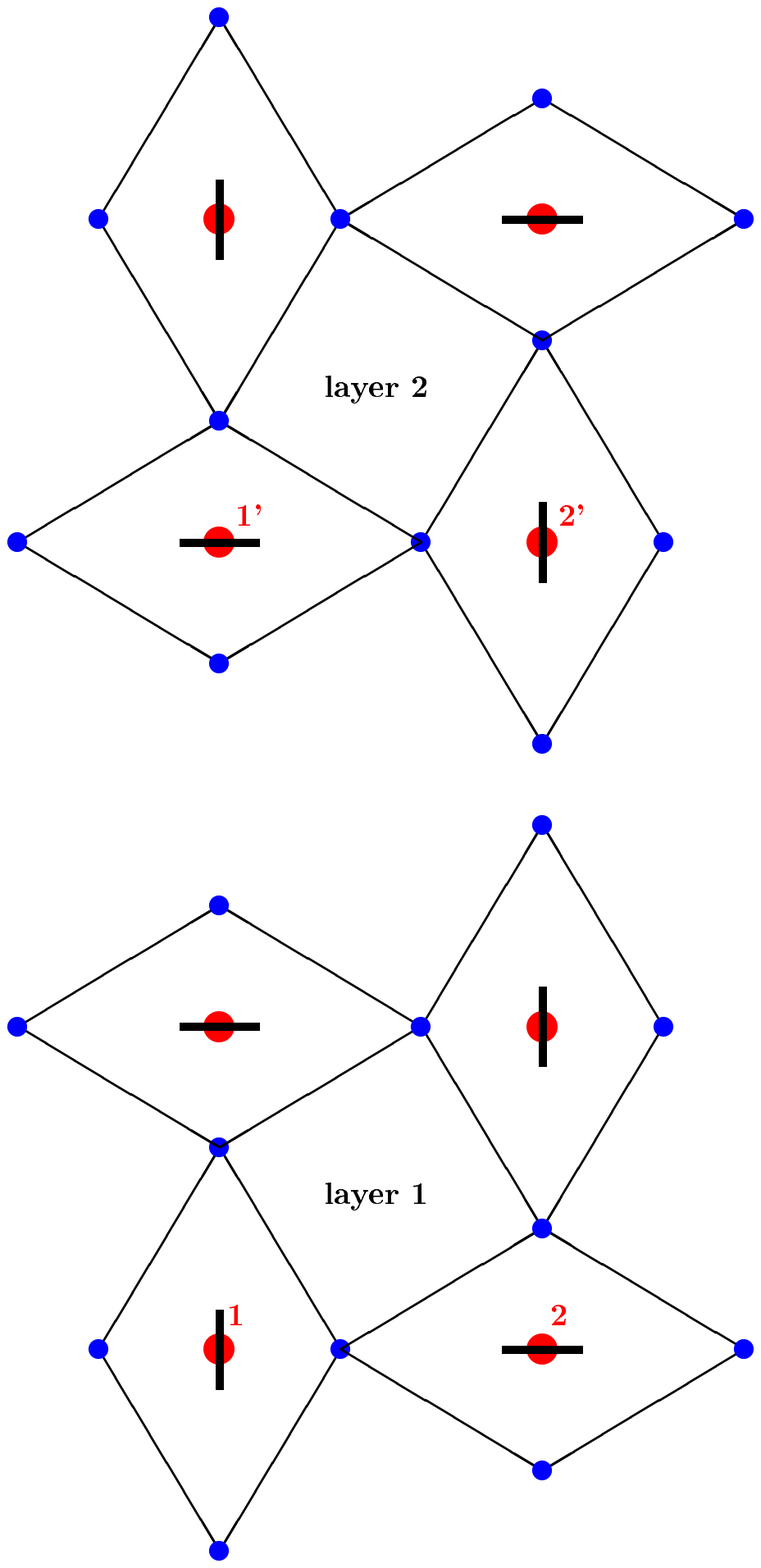}
\end{center}
\vspace{-4.0cm}
\caption{%
Schematic view of JT distortions used for Hartree-Fock computations in
the zero-temperature phase of LaVO$_3$.
Here over the first layer 1 the layer 2 is stacked and the orbitals
$\{yz,zx\}$ form $G$-AO order, i.e., alternate along the $c$ axis.
The meaning of other symbols is the same as in Fig. 2.
}
\label{fig3}
\end{figure}

The purpose of this paper is to investigate the spin-orbital order in
vanadium perovskites within the multiband $d$-$p$ model, i.e., to go
beyond the usually used picture of a Mott insulator with $S=1$ spins
and $t_{2g}$ orbital degrees of freedom or effective degenerate Hubbard
model of $t_{2g}$ electrons. The $d-p$ model includes non-zero on-site
Coulomb interactions defined both on oxygen and on transition metal ions
and takes into account the possibility of finite self-doping, explained
below and applied before to ruthenium, iridium, and titanium oxides
\cite{Ros15,Ros16,Ros17}.
The $d-p$ model was developed in these papers into a realistic method,
capable of computationally cheap and fast realistic investigation of
the electronic structure of complex transition metal oxides.

Up to now, the on-site Coulomb interactions on oxygen ions are being
neglected in the majority of papers (as a simplification --- to reduce
the computational effort). However, when Coulomb repulsion elements on
oxygens are neglected, the true $d-p$ model parameters are replaced by
effective parameters. In particular, the ''effective'' Hubbard
repulsion on vanadium ions $U_d$ is smaller by about 50\% than the
``true'' $U_d$ repulsion \cite{Ros17}. Also the so-called
\textit{self-doping}  \cite{Ros15,Ros16,Ros17}, see below, is neglected
in traditional effective $3d$-electron models where one assumes that a
cation (for example La in LaVO$_3$ or Y in YVO$_3$) behaves according
to the {\em idealized ionic model}, i.e., donates {\em all} valence
electrons into a VO$_3$ unit (for La these are: two $4s$ electrons and
one $3d$ valence electron). However, in reality, this charge transfer 
is smaller --- it is not exactly 3 but $(3-x)$ instead.
Strictly speaking, we mean by this statement that the occupation number
of valence electrons on La as obtained say by Mulliken (or Bader)
population analysis (during a parallel \textit{ab-initio} computation)
will amount to some finite value of $x>0$. This redistribution of
electron charge is called here \textit{self-doping}.

In the present
investigation we use up-to-date estimations of crystal-field
splittings, spin-orbit interaction at vanadium ions and JT distortions.
The model is used to study possible types of order, and to establish
the easy spin-axis. We also extracted from our computations HOMO-LUMO
gaps which can serve as an estimation of the band gap.

The paper is organized as follows. We define the model and its
parameters in Sec. \ref{sec:model}. The numerical method and its
caveats are addressed in Sec. \ref{sec:num}. The results are
presented and discussed in Sec. \ref{sec:res}. In Sec.
\ref{sec:summa} we present the main conclusions and a short summary.

\section{Hamiltonian}
\label{sec:model}

We introduce the multi-band $d-p$ Hamiltonian for VO$_3$
three-dimensional (3D) cluster which includes five $3d$ orbitals at
each vanadium ion and three $2p$ orbitals at each oxygen ion,
\begin{equation}
{\cal H}= H_{\rm kin}+H_{\rm so}+ H_{\rm diag} +H_{\rm int}.
\label{model}
\end{equation}
where $H_{\rm kin}$ stands for the kinetic energy,
$H_{\rm so}$ for spin-orbit coupling,
$H_{\rm diag}$ for the diagonal part of kinetic energy
(also including local crystal-field splittings), and
$H_{\rm int}$ for the intraatomic Coulomb interactions.
Optionally one can add JT part $H_{\rm JT}$ and this will be
discussed in Sec. \ref{sec:JT}. The cluster geometry and precise
forms of different terms are standard;
for the detailed formulas see Refs. \cite{Ros15,Ros16}.

The kinetic part of the Hamiltonian is:
\begin{equation}
H_{\rm kin}=\sum_{\{i\mu; j,\nu\},\sigma}\left(t_{i,\mu;j,\nu}
 c^{\dagger}_{i,\mu,\sigma}c_{j,\nu,\sigma}^{} + H.c.\right),
\end{equation}
where we employ a general notation, with $c_{j,\nu,\sigma}^{\dagger}$
standing for the creation of an electron at site $j$ in an orbital
$\nu$ with up or down spin, $\sigma=\uparrow,\downarrow$.
The model includes all five $3d$ orbital states
$\nu\in\{xy,yz,zx,x^2-y^2,3z^2-r^2\}$, and three $2p$ oxygen orbital
states $\nu\in\{p_x,p_y,p_z\}$. Alternatively, i.e., when choosing a
more intuitive notation, we can write $d_{j,\nu,\sigma}^{\dagger}$ for
$d$ orbitals, while $p_{j,\nu,\sigma}^{\dagger}$ for $p$ orbitals.
The matrix $t_{i,\mu j,\nu}$ is assumed to be non-zero only for
nearest neighbor vanadium-oxygen $d-p$ pairs, and for nearest
neighbor oxygen-oxygen $p-p$ pairs. The next-nearest hopping elements
are neglected. (The nonzero $t_{i,\mu; j,\nu}$ elements are listed in
the Appendix of Ref. \cite{Ros15}; we use the Slater notation
\cite{Sla54}).
As a side remark we recall that models taking into account only three
$t_{2g}$ orbitals and neglecting remaining two $e_g$ orbitals are
not accurate enough \cite{Yan11}.

The spin-orbit part,
$H_{\rm so}=\zeta\sum_i\textbf{L}_i\cdot\textbf{S}_i$,
is a one-particle operator (scalar product of angular momentum and spin
operators at site $i$), and therefore it can be represented in the form
similar to the kinetic part $H_{\rm kin}$ \cite{Miz96,Pol12,Mat13,Du13},
\begin{equation}
H_{\rm so}= \sum_i\left\{\sum_{\mu\neq\nu;\sigma,\sigma'}
t^{so}_{\mu,\sigma;\nu,\sigma'}
d^{\dagger}_{i,\mu,\sigma}d_{i,\nu,\sigma'}^{} + \mathrm{H.c.}\right\},
\label{so-part}
\end{equation}
with $t^{so}_{\mu,\sigma;\nu,\sigma'}$ elements restricted to single
vanadium sites. They all depend on spin-orbit strength $\zeta$
($\zeta=0.026$ eV; this value was adopted from Ref. \cite{Dai08})
which is weak but it can have influence on the preferred spin
direction. For detailed formula and tables listing
$t_{i\;\nu,\sigma';\mu,\sigma}$ elements, see Refs. \cite{Ros15,Pol12}.

The diagonal part $H_{\rm diag}$ depends only on electron number
operators. It takes into account the effects of local crystal fields
and the difference of reference orbital energies (here we employ the
electron notation),
\begin{equation}
\Delta=\varepsilon_d-\varepsilon_p,
\label{Delta}
\end{equation}
between $d$ and $p$ orbitals (for bare orbital energies).
We can fix the reference energy $\varepsilon_d=0$ for $d$ orbitals to
zero and use only $\Delta=-\varepsilon_p$ as a parameter, thus we write
\begin{eqnarray}
H_{\rm diag}=
\sum_{i;\mu=x,y,z;\sigma}
\varepsilon_p^{} p^\dagger_{i,\mu,\sigma}p_{i,\mu,\sigma}^{} \nonumber \\
+\sum_{m;\mu=xy,yz,... ;\sigma}
f^{cr}_{\mu,\sigma} d^\dagger_{m,\mu,\sigma}d_{m,\mu,\sigma}^{}.
\end{eqnarray}
The first sum is restricted to oxygen sites, while the second one runs
over vanadium sites. The crystal-field splitting strength vector
($f^{cr}_{\mu,\sigma}$) describes the splitting within $t_{2g}$ levels.
For example, in YVO$_3$ the $xy$ orbital is lowered by $\sim 0.017$ eV
(according to Ref. \cite{Ots06}). At the same time, the $\{yz,zx\}$
doublet is also split (this was discussed in some papers, most clearly
in Refs. \cite{Jo03,Hor08}) in accordance with local JT distortion of
particular VO$_6$ octahedron. \textit{We assume ad-hock} that either
$yz$ is lower than $zx$ orbital which should correspond to O$_4$ square
(in $ab$ plane) when distorted from ideal square into elongated along
$y$-direction rhombus, or the opposite: $zx$ is lower than $yz$ orbital
which should correspond to O$_4$ distorted into elongated along
$x$-direction rhombus (compare Figs. 1 and 2). This splitting value
should be $0.1-0.2$ eV what is an educated guess
(compare with the estimation from \mbox{Ref. \cite{Ots06}}).

The distance between $t_{2g}$ levels and $e_g$ levels is large,
$1.5-2.0$ eV \cite{Mos09,Mos10,Reu12}). We do not take into account a
possible splitting within $e_g$ levels as from our previous experience
with transition metal perovskites we do not expect it to be an
important factor.

The on-site Coulomb interactions $H_{\rm int}(d)$ for $d$ orbitals take
the form of a degenerate Hubbard model \cite{Ole83},
\begin{eqnarray}
H_{\rm int}(d)&=&
\sum_{m,\mu<\nu}\left(U_d-\frac{5}{2} J^d_{\mu\nu}\right)
n_{m\mu} n_{m\nu}\nonumber\\
&+&U_d \sum_{m\mu}  n_{m\mu\uparrow} n_{m\mu\downarrow}
-2 \sum_{m,\mu<\nu}J^d_{\mu\nu}\,\vec{S}_{m\mu}\cdot\vec{S}_{m\nu}
\nonumber\\
&+&  \sum_{m,\mu\neq\nu} J^d_{\mu\nu}\,
d^\dagger_{m\mu\uparrow} d^\dagger_{m\mu\downarrow}
d_{m\nu\downarrow}^{}d_{m\nu\uparrow}^{}.
\label{hubbard2-intra}
\end{eqnarray}
where $n_{m\mu}=\sum_{\sigma}n_{m\mu\sigma}$ is the electron density
operator in orbital $\mu$, $\{\mu,\nu\}$ enumerate different $d$
orbitals, and $J_{d,\mu\nu}$ is the non-trivial tensor of on-site
interorbital exchange (Hund's) elements for $d$ orbitals;
$J_{d,\mu\nu}$ has different entries for the $\{\mu,\nu\}$ pairs
corresponding to two $t_{2g}$ orbitals ($J_{\rm H}^t$), and for a pair
of two $e_g$ orbitals ($J_{\rm H}^e$), and still different for the case
of cross-symmetry terms \cite{Ole05,Hor07}; all these elements are
included and we assume the Racah parameters: $B=0.1$ eV and $C=4B$.

The local Coulomb interactions $H_{\rm int}(p)$
at oxygen sites (for $2p$ orbitals) are analogous,
\begin{eqnarray}
H_{\rm int}(p)&=&
\sum_{i,\mu<\nu,\sigma} \left(U_p-\frac{5}{2} J^p_{\rm H}\right)
n_{i\mu} n_{i\nu} \nonumber\\
&+&U_p \sum_{i\mu}  n_{i\mu\uparrow} n_{i\mu\downarrow}
-2J^p_{\rm H}\sum_{i,\mu<\nu} \vec{S}_{i\mu}\cdot\vec{S}_{i\nu}
\nonumber\\
&+& J^p_{\rm H}\sum_{i,\mu\neq\nu}
p^\dagger_{i\mu\uparrow}p^\dagger_{i\mu\downarrow}
p_{i\nu\downarrow}^{}p_{i\nu\uparrow}^{},
\label{hubbard3-intra}
\end{eqnarray}
where the intraatomic Coulomb repulsion is denoted as $U_p$ and
all off-diagonal elements of the tensor $J^p_{\mu\nu}$ are equal
(as they connect the orbitals of the same symmetry), i.e.,
$J^p_{\mu\nu}\equiv J^p_{\rm H}$. (Up to now, as already mentioned
above, $H_{\rm int}(p)$ was neglected in the majority of studies, i.e.,
for simplicity it was being assumed that $U_p=J^p_{\rm H}=0$.)

In the following we use the parameters $U_d$, $J^d_{\mu\nu}$, $U_p$,
and $J_{\rm H}^p$ similar to those used before for titanium oxides
\cite{Ros16,Ros17}; for the hopping integrals we follow the studies by
Mizokawa and Fujimori \cite{Miz95,Miz96}. The value $U_p\sim 4.0$ eV
was previously used in copper oxides \cite{Hybe92,Sing13} but in
addition in some test computations we considered a larger value $U_p=6$
eV. (This choice, i.e., $U_p=6$ eV
is advocated and reasonably explained in Refs. \cite{Sing13,park12}.)
Concerning the parameter $\Delta$ an educated guess is necessary as no
information for the vanadium perovskites is available. However, we have
found before that in titanium oxides $\Delta=6.5$ eV is reasonable
\cite{Ros16,Ros17}. Here for vanadium oxides a smaller value should be
more appropriate. Old-fashioned computations, such as those reported
in the classical textbook of Harrison \cite{Har05} and shown in tables
therein suggest a value lower by 1.5 eV (i.e., $\Delta=5.0$ eV);
a still lower value of 4.0 eV was suggested by Bocquet \textit{et al.}
and Imada \textit{et al.} \cite{Ima98} (note that in these papers the
parameter $U_p$ enters only indirectly). We have tried all values in
the range $4.0<\Delta<6.5$ eV and found that the most interesting and
sensible physical results could be obtained for $\Delta=5.0$ eV.

\begin{table}[t!]
\caption{Parameters of the multi-band model (\ref{model}) (all in eV)
used in the calculations. For the hopping integrals we adopt the values
from Refs. \cite{Miz95,Miz96}, i.e.,
$(pd\sigma)\;(pd\pi)\;(pp\sigma),\;(pp\pi)=-2.2,\,1.1,\,0.6,\,-0.15$ eV
which correspond to V$-$O distances of 2.0 \AA $\,\,$ (we use Slater
notation \cite{Sla54}). The charge transfer energy
(defined for bare levels) is taken as $\Delta=5.0$ eV.}
\begin{ruledtabular}
\begin{tabular}{cccccc}
$\zeta $ & $U_d$ & $J_{\rm H}^t$ & $J_{\rm H}^e$ & $U_p$ & $J_{\rm H}^p$   \\
  \hline
0.026 & 8.0  &  0.8  &  0.9  &  4.4  &   0.8   \\
\end{tabular}
\end{ruledtabular}
\label{tab:para}
\end{table}

Our reference system is LaVO$_3$ where the total electron number in the
$d-p$ subsystem is $N_e=17+3=20$ per one VO$_3$ unit provided we assume
an ideal ionic model with no self-doping ($x=0$), i.e., all three La
valence electrons are transferred to VO$_3$ unit. Another possibility
is when the self-doping is finite: we consider $x=0.5$
(then the cation La donates not 3 but rather on average 3-0.5=2.5
electrons and $N_e=20-x=19.5$); or the extreme $N_e=19$ when the
self-doping is $x=1.0$. Note that in the following for our computations
we use only certain discrete numbers for $x$ as the studied cluster is
finite and the total electron number must be an even integer; moreover
the total electron number should hit some magic number so that the
ground state wave function of the studied \textit{small cluster} is
close-shell and not an open-shell.

The problem how to fix $x$ is a difficult question. If one wants to
be sure what is a precise value of $x$, then the best way would be to
perform independent, auxiliary \textit{ab-initio} or local density
approximation with Coulomb interaction $U$ (LDA+$U$) computations and
extract the electronic population on the cation $R$ (in $R$VO$_3$)
analogously like it was done in Ref. \cite{Ros17}. This is however
rather expensive. Without such auxiliary \textit{ab-initio}
computations one is left with speculations. It seems that for the case
of La or Y cation a safe guess is that $x\in[0.0,0.5]$, i.e.,
all three, or almost all three $5d^16s^2$ valence electrons are
transferred to the vanadium octahedron.

\section{Numerical studies}
\label{sec:num}

\subsection{Computational problems concerning the Jahn-Teller Hamiltonian}
\label{sec:JT}

The important part of the electronic Hamiltonian in perovskites, namely
the influence of JT distortions on the electronic structure rarely can
be treated in a satisfactory way during the computations.
Let us explain what we mean by this statement. An effective Hamiltonian
which describes cooperative JT lattice distortions for octahedra in the
vanadium perovskites can be assumed in the complicated form which is
quadratic in JT distortions and contains in addition the terms
$\propto d_{i\nu\sigma}^\dagger d_{i\mu\sigma}^{}$ coupled linearly
with JT distortions, for details and explicit (quite complicated)
formula, see for instance Ref. \cite{Mul10}.
JT distortions $\{Q_i\}$ ($i=1,\dots,6$) (used notation is the same as
in Ref. \cite{Kan60}) can be treated as quasi-classical continuous
variables. There should be appended (to all $Q's$) an additional
(extra) subscript $m$ to distinguish between different octahedra which
could have (in principle) different, one from another, distortions.
Let us remind that (see Ref. \cite{Kan60}) $\{Q_4,Q_5,Q_6\}$ modes
cause tilting (rotations) of the VO$_6$ octahedron. The $Q_2$ mode
causes distortion of squares formed by four oxygens (in $ab$ plane a
square undergoes distortion into an elongated rhombus), while the
$Q_3$ mode causes differences in apical vanadium-oxygens
bond lengths (tetrahedral distortion).

In the course of normal computations (when looking for ground state
energy minimum) the search for energy minimum due to electronic degrees
of freedom must be supplemented with an extra search for the optimal
values of continuous classical degrees of freedom ($Q_i$-modes). Then
the Hamiltonian becomes intractable, even so for very small clusters,
even so if the cooperative pattern of JT distortions is explicitly
assumed. Let us remark that assuming cooperative and static pattern of
JT distortions (with a certain amount of symmetry) would mean that
instead of $Q_{2m},Q_{3m},\dots$ (a lot of separate sets of
$Q_{2m},Q_{3m},...$, one set for each individual octahedron $m$) one
can consider a single set of $|Q_2|,|Q_3|,\dots$ and the dependence on
the octahedron number $m$ within the lattice is realized through
alternating plus/minus signs to individual $Q$'s and changing them
according to the assumed global symmetry of the static-cooperative JT
distortions. Anyway, even with this great simplification there are at
least five extra $\{Q_i\}$ variables which makes looking for ground
state energy minimum during HF iterations extra expensive.

To overcame this difficulty most often a semiempirical treatment of JT
terms is used: namely one assumes an explicit form and the magnitudes
of the lattice distortions, usually suggested by the experiment. Thus
the distorted lattice is frozen and we take this as an experimental
fact (and do not ask any more about the origin of these distortions).
Then computations become more feasible. The $Q_{2m},Q_{3m},\dots$ modes
and the JT Hamiltonian do not enter computations anymore --- their
only role was to deform the lattice and to change V$-$O distances.
Instead, one collects all V$-$O and O$-$O bond lengths (as suggested
by experiment) and because of modified bond lengths one modifies the
matrix of kinetic hopping parameters. In this respect quite popular is
the Harrison scaling \cite{Har05} when the difference in V$-$O bond
lengths (versus some reference bond lengths, for example those in
hypothetical undistorted crystal of ideal cubic symmetry) causes
renormalization of the hopping elements. The second important
consequence of changed V$-$O distances is creation of local crystal
fields acting upon central V-ions: these will split $yz/zx$ doublets
as already discussed above for $H_{\rm diag}$ and $f^{cr}_{\mu,\sigma}$.

To simplify the numerical effort, we performed exactly such
computations but only for scenario shown in Fig. 2, i.e., only $Q_2$
distortions were included, while the $Q_4$ distortions were neglected.
This choice is purely pragmatic: non-zero $Q_4$'s, $Q_5$'s, and $Q_6$'s
significantly increase computational effort by drastically lowering the
symmetry and therefore increasing the complexity of kinetic hopping
matrix. We emphasize that the $d-p$ model is definitely not an
\textit{ab-initio} approach thus it can account only for a qualitative
description of generic physical properties; one should not expect that
all the physical details will be described properly. Therefore certain
simplifications in modeling are not a capital offense. In this respect
one can still ask if indeed octahedral tilting and finite $Q_4$
distortions are mandatory for spin-orbital order to emerge.
Numerous experimental and theoretical papers addressed directly and
indirectly these questions:
(i) quoting Ref. \cite{Yan07} where the proof was given than V$-$O$-$V
angles deviating strongly from 90\textdegree~are not primary a driving
force stabilizing $C$-type orbital order in vanadates, or that
(ii) orbital fluctuations (at zero temperature) are not strong but in
fact almost suppressed \cite{Ray07}.
For a more general discussion of these problems see Ref. \cite{Fuj10}.
We suggest that for the description of the onset of spin-and-orbital
order, our simplified scenarios with local crystal fields and with
geometries depicted in Figs. 2 and 3 are quite enough and that the
apical axes non-zero tiltings influence only the distances between the
true HF ground state and other (higher in energy) stable HF states.

To summarize, and at the same time to give an explicit example: In
YVO$_3$ we studied the zero-temperature geometry as shown in Fig. 2
with repeating layer 1 (along the $c$ axis): the V$-$O bond lengths
were set as 2.042, 1.99, and 1.99 \AA $\,\,$ \cite{Bla01,Kha04} for a
long, a short, and an apical bond, respectively. The Slater integrals
were scalled following the Harrison's rules \cite{Har05} to fit the
experimental V$-$O bond lengths. The changes of O$-$O bond lengths
caused by JT distortions were neglected (they are expected to be small
and less important). On top of it the values of local crystal field
splitting of $yz/zx$ doublet were assumed to be $\pm 0.1$ eV.

\subsection{Unrestricted Hartree-Fock computations}

We use the unrestricted HF approximation (with a single determinant
wave function) to investigate the model (\ref{model}).
The technical implementation is the same as that described in Refs.
\cite{Ave13,Miz95,Miz96,Sug13,Ros15,Ros16} featuring the averages
$\langle d^\dagger_{m,\mu,\uparrow}d_{m,\mu,\uparrow}^{}\rangle$
and  $\langle p^\dagger_{i,\mu,\uparrow} p_{i,\mu,\uparrow}^{}\rangle$
(in the HF Hamiltonian) which can be treated as order parameters.
At the beginning some initial values (a guess) have to be assigned to
them. During HF iterations the order parameters are recalculated
self-consistently until convergence.
If in the course of computations all the averages
$\langle d^\dagger_{m,\mu,\uparrow}d_{m,\mu,\uparrow}^{}\rangle,\dots$
would be treated as independent, convergence (if any) would indeed be
too slow. Therefore the common strategy is to employ explicit type of
symmetry of the order in the ground state
(which lowers the number of order parameters) and to perform HF
iterations strictly under this assumption. During present computations
the chosen scenarios for the ground state symmetry were those with
either:
(i) orbital order of $G$-type, or $C$-type, or absent;
(ii) spin order $G$-AF, or $C$-AF, or absent;
(iii) $x$ or $z$ easy magnetization axis.
One remark: the hypothetical ground state symmetries which would
violate Goodenough-Kanamori rules \cite{Goode} were also considered
(these are: ground state with $C$-AF spin order and $C$-type orbital
order and also ground state with $G$-AF spin order and $G$-type orbital
order; during computations we found such states to be locally stable in
unrestricted HF for some parameters, but they never became true ground
state).

Within each of the above scenarios the number of independent order
parameters is lowered but still it is large enough so that the HF
convergence is rather poor. This was caused mainly by not imposing any
restrictions on order parameters associated with oxygens (no orbital
equivalence by symmetry, no assumption on oxygens magnetic properties)
and not imposing any symmetry restriction on order parameters
associated with vanadium $e_g$ orbitals. We found that imposing any of
such restrictions is quite risky as any symmetries and orbital
equivalences as could be \textit{a priori} assumed, in fact turn out to
be too restrictive and only approximate ones. This happens at least for
scenarios shown in Figs. 2 and 3. For computations and a quick scan of
the phase diagram we used $2\times 2\times 4$ cluster. (A single HF run
on an ordinary desktop can be done in about 10 minutes; bigger
$4\times 4\times 4$ clusters require from several hours up to one day).

The simplified and popular remedy for poor HF convergence is the
so-called dumping technique. Better remedy is the technique known in
quantum chemistry and called level shifting \cite{Sou73}. It is based
on replacing the true HF Hamiltonian by a different Hamiltonian ---
the one with the identical eigenvectors (one particle eigenfunctions)
as the original Hamiltonian and with identical \textit{occupied}
eigenenergies. The original eigenenergies of virtual states are however
uniformly shifted upwards by a fixed constant value. Thus if we apply
the shift say by 5.0 eV, then the gap between the highest occupied
molecular orbital and the lowest unoccupied molecular orbital
(HOMO-LUMO gap) we obtain will be artificially enlarged exactly by
5.0 eV. (So, it has to be corrected by subtracting from the obtained
HOMO-LUMO gap the fixed value of 5.0 eV).

When applying virtual level shifting we can obtain some additional
information. Namely when the HOMO-LUMO splitting (after correcting for
the shift) is negative, then the single-determinant HF ground state we
obtained is not correct (this assumes that sufficient number of
different HF starting conditions was tried). One possibility is that
the true ground state is conducting, another is that a
single-determinant HF wave function breaks down due to very strong
electronic correlations and multi-configuration HF method is required.

\section{Results and discussion}
\label{sec:res}

\subsection{Zero-temperature ground state in LaVO$_3$}

The symmetry of LaVO$_3$ at zero temperature is monoclinic
\cite{Zub76,Ren03} which should correspond to $G$-type orbital order
(which is induced, or to say it directly, is enforced by cooperative
crystal field splittings of $yz/zx$ doublets).
The bond lengths at zero temperature were
difficult to find in the literature --- following Ref. \cite{Saw96}
we took 2.04 \AA $\,\,$/1.98 \AA $\,\,$ for long/short V-O distances
within $ab$ plane and 1.98 \AA $\,\,$ for apical V-O bonds. With this
choice we assumed the following local crystal field values: 1.8 eV as
the distance between $t_{2g}$ levels and $e_g$ levels, and an ad-hock
choice: $\pm 0.10$ eV as splitting between $yz/zx$ orbitals, also $xy$
orbital energy is lowered
(due to tetragonal distortion of V$-$O apical bonds) by 0.1 eV.

The experimentally found spin-order is $C$-AF with average magnetic
moment $|\langle m\rangle|\in(0.6,0.7)$ and easy magnetization axis $c$
\cite{Zub76,Tep00}. The estimations of band gaps are in between 1.1 and
1.8 eV \cite{Tsv04,Mos09,Mos10,Kum17}; the most popular value is 1.1 eV.
Below in Table II we collected the obtained results for hypothetical
self-dopings $x=0,\,0.5,\,1.0$ (we remind that we do not know which
one of these values is closest to the true one). Some comments about
the legend in Table II: the indices $m=1$ and 2 in
$\langle n_{1,xy,\uparrow}\rangle,\dots$, etc., stand for two
nonequivalent vanadium ions, see Fig. 2. $E_{\rm HF}$ is the HF energy
per one VO$_3$ unit, $G$ is the HOMO-LUMO gap, $\langle m\rangle$ is an
average magnetic moment per V ion (when expressed in $\mu_B$ it should
be twice larger). In Table II the spin-order type with an easy
magnetization direction is indicated, and finally $x$ is self-doping
level which was fixed during computations.

Now we summarize the results obtained for different electronic filling
of VO$_3$ octahedra (self-doping). We start with self-doping $x=0$
which stands for an ideal ionic model. The best HF ground state
reproduces the experimental spin-orbital order found in LaVO$_3$,
see the $x=0$ column of Table II. However, the next candidate for the
HF ground state with $C$-AF spin order parallel with the $x$ axis (see
Fig. 2), which corresponds to (1,1,0) crystallographic direction and
is only by 0.3 meV energetically higher (not shown). Note that when
spin-orbit interaction is neglected the change is here insignificant:
instead of 0.3 meV we obtained 0.2 meV energy difference.
(A general discussion of the role played by spin-orbit interaction in
the vanadium perovskites was presented in Refs.
\cite{Hor03,Yan04,Zho07}).

For spin order along the $z$ axis site
$m=1$ corresponds to magnetization $m_1\simeq 0.99$ and site $m=2$ to
$m_2\simeq -0.99$. We observe that when quantum fluctuations are absent
as in our calculation, the magnetization is somewhat reduced due to
minority-spin electron density in the occupied $t_{2g}$ orbitals, while
this reduction is almost fully compensated by majority-spin electron
density in the empty $t_{2g}$ and two $e_g$ orbitals. In this way we
arrive at $|\langle m\rangle|\simeq 0.99$ which results from electron
delocalization by $d-p$ hybridization. It is remarkable that total
electron density in $e_g$ orbitals is close to 0.40 which definitely
shows that $e_g$ orbitals contribute to the electronic structure.
What concerns the average occupation of $2p$ electrons:
(i) on oxygens aligned along the $x$ axis (see Fig. 2) it is 5.80 with
average moments either 0.0 or $\pm 0.01$ (changing not randomly but in
a regular way);
(ii) for oxygens aligned along the $y$ axis the corresponding numbers
are 5.86 for the charge and 0.0 or $\pm 0.01$ for the moments;
(iii) for oxygens aligned along the $z$ axis  (this coincides with the
crystallographic $c$ direction) the occupation is 5.80 and no moments
are found. The next HF stable state is by 1.3 meV higher than the true
ground state --- it has $G$-AF spin order parallel to the $z$-axis.
Note that this state violates Goodenough-Kanamori rules \cite{Goode}.

\begin{table}[t!]
  \caption{Spin and orbital order and the
    electron occupations on vanadium ions for the zero
  temperature HF ground states of LaVO$_3$ and YVO$_3$.
Subscripts $x$  and $z$ in $C$-AF$_x$ and $C$-AF$_z$ denote
the axis of easy-magnetization (compare Fig. 2).
}
\begin{ruledtabular}
  \begin{tabular}{lccc}
 $x$             &    0.0        &    0.5        &   1.0             \\  \hline
\multicolumn{4}{c}{ LaVO$_3$ (monoclinic) } \\
\mbox{orbital order}  &  $G$-AO   &  $G$-AO    &  \mbox{none}    \\
\mbox{spin order}  &  $C$-AF$_z$   &  $C$-AF$_z$    &  $C$-AF$_x$    \\
$E_{\rm HF}$ (eV)       & 32.555   & 26.573  & 20.673 \\
$G$   (eV)              &  3.92    &  1.99   &     2.89   \\
$|\langle m\rangle|$    &  0.99    & 0.77    &  0.52    \\   \hline
\multicolumn{4}{c}{ YVO$_3$ (orthorhombic) } \\
\mbox{orbital order}  &  $C$-AO   &  $C$-AO    &  \mbox{none}    \\
\mbox{spin order}  &  $G$-AF$_z$   &  $G$-AF$_z$    &  $C$-AF$_x$    \\
$E_{\rm HF}$ (eV)       & 32.734   & 26.802  & 20.935 \\
$G$   (eV)              &  3.99    &  2.02   &     2.86   \\
$|\langle m\rangle|$    &  0.99    & 0.77    &  0.52    \\   \hline  \hline
\multicolumn{4}{c}{ \mbox{electron occupations} both for LaVO$_3$ and YVO$_3$ } \\
$\langle n_{1,xy,\uparrow} \rangle   $ &1.00    & 0.86  & 0.54       \\
$\langle n_{1,xy,\downarrow} \rangle $  &0.05   & 0.06  &  0.54          \\
$\langle n_{1,yz,\uparrow} \rangle   $ &1.00    & 0.60  &0.12       \\
$\langle n_{1,yz,\downarrow} \rangle $  &0.04   & 0.06  &0.12         \\
$\langle n_{1,zx,\uparrow} \rangle   $ &0.09    & 0.23 &  0.12          \\
$\langle n_{1,zx,\downarrow} \rangle $  &0.06    & 0.08 &  0.12           \\
$\langle n_{1,x^2-y^2,\uparrow} \rangle   $ &0.10    &  0.12  &0.13        \\
$\langle n_{1,x^2-y^2,\downarrow} \rangle $ &0.08    &   0.10  &0.13        \\
$\langle n_{1,3z^2-r^2,\uparrow}\rangle   $ &0.13    &  0.16 &0.18          \\
$\langle n_{1,3z^2-r^2,\downarrow}\rangle $ & 0.10    &  0.13  &0.18        \\
\hline
$\langle n_{2,xy,\uparrow} \rangle   $ &0.05    &  0.06   &0.54      \\
$\langle n_{2,xy,\downarrow} \rangle $ &1.00    &  0.86   &0.54         \\
$\langle n_{2,yz,\uparrow} \rangle   $ &0.06    & 0.08   &0.12      \\
$\langle n_{2,yz,\downarrow} \rangle $ &0.09    &  0.23  &0.12         \\
$\langle n_{2,zx,\uparrow} \rangle   $ &0.04    &  0.06  & 0.12         \\
$\langle n_{2,zx,\downarrow} \rangle $ & 1.00    &   0.60 &0.12         \\
$\langle n_{2,x^2-y^2,\uparrow} \rangle   $ & 0.08     & 0.10 & 0.13         \\
$\langle n_{2,x^2-y^2,\downarrow} \rangle $ & 0.10   &   0.12  &0.13       \\
$\langle n_{2,3z^2-r^2,\uparrow}\rangle   $ &  0.10    & 0.13 & 0.18          \\
$\langle n_{2,3z^2-r^2,\downarrow}\rangle $ &0.13    &  0.16  &0.18        \\
\end{tabular}
\end{ruledtabular}
\end{table}

The states with different spin order are almost degenerate. Most
probably a more complicated geometry featuring sizable octahedral axes
tilting should account for bigger differences, such as those reported
in Ref. \cite{Miz99}. Summarizing, spin-orbital order for $x=0$ is
\textit{ideally reproduced} with respect to present paradigm of
spin-orbital order in vanadates \cite{Bla01} but average spin and
band-gap we obtained do not agree too well with the experimental values.

Consider now doping $x=0.5$: The best HF ground state we obtained here
also reproduces correctly the experimental spin-orbital order found in
LaVO$_3$, see the third column of Table II. The next candidate for the
ground state is the one with $C$-AF spin order but this time aligned
along the $x$ axis (see Fig. 2).
Actually, it is by 2.0 meV higher; note that the
spin-orbit interaction is here more important and responsible for so
large energy difference; when this interaction is absent one finds
instead the energy difference of 0.6 meV. Magnetization of
$|\langle m\rangle|\simeq 0.77$ corresponds better to the experiment
--- one finds here definitely weaker magnetization contributions from
two occupied $t_{2g}$ orbitals but a larger magnetization in the third
$t_{2g}$ orbital. Altogether, electron density in $t_{2g}$ orbitals is
lower than that at $x=0$, but at the same time the $e_g$-electron density
(but not magnetization) is somewhat enhanced.

The oxygen electron occupations indicate charge delocalization by $d-p$
hybridization in presence of spin-orbit coupling:
(i) for oxygens along the $x$ axis electron density is 5.73 while
magnetic moments are $\pm 0.01$;
(ii) for oxygens along the $y$ axis electron densities of 5.57 and
5.59 are accompanied by $\pm 0.01$ moments (arranged with a suitable
regularity, both spins and the tiny charge modulation);
(iii) for oxygens along the $z$ axis  (this coincides with the
crystallographic $c$ axis) occupations are 5.81 with zero moments,
or 5.76 with $\pm 0.02$ moments --- again both spins and tiny charge
density wave are arranged with a suitable regularity.
Note that average spin value of 0.77 and the HOMO-LUMO gap of 1.99 eV
fit rather well to the experimental results. Thus we suggest that for
LaVO$_3$ the self-doping is $x\approx 0.5$ and that the entries from
the third column in Table II are a rather faithful description of the
experimental situation.

At this point we would like to make a short digression and explain in
a more transparent way why a weak ($x=0.5$) self-doping effect is
important in LaVO$_3$. It is true that the spin and orbital order for
$x=0$ and $x=0.5$ are qualitatively identical. However, the average
magnetization (per V ion) is $\sim 1.0$ for the pure ionic model $x=0$
and this is unrealistic. At the same time for $x=0.5$ the average
computed magnetization value drops to 0.77 --- this is more realistic
and quite close to the experimental value. We conclude that self-doping
reduces the order parameter by including the covalency effect.

There is also a second argument: the band gap we computed for $x=0.5$
is much closer to the experimental value that band gap we computed for
$x=0$. It is well known that Hartree-Fock computations tend to
overestimate band gaps. And indeed, for $x=0.5$ we obtained
$G\approx 2.0$ eV, while the experimental values indicate $1.1<G<1.8$
eV. However our overestimation of the gap (probably by $\approx$30\%)
is not that severe as in case of $x=0$ where we obtain
$G\approx 4.0$ eV. These two facts clearly suggest that including
weak self-doping effect is important for realistic modelling of the
vanadium perovskites.

For large self-doping $x=1.0$ orbital order disappears. Only $xy$
orbitals are occupied by approximately one electron, while all the
densities in all other ($t_{2g}$ and $e_g$) orbitals are close to
0.25, with somewhat enhanced density of 0.36 in $3z^2-r^2$ orbitals.
Note that this large density follows from the delocalization of $2p$
electrons from oxygen ions. The ground state has solely spin $C$-AF
order with $x$ easy axis of magnetization. This state contradicts
experimental observations and excludes so high self-doping level.
No entry in the last column
of Table II provides the direct evidence that the spins align indeed
along the $x$ axis. To supplement this information we must make
another digression. Thus we note that at the $m$-th vanadium ion,
$\langle d^\dagger_{m,\mu,\uparrow}d_{m,\mu,\uparrow}^{}\rangle=
\langle d^\dagger_{m,\mu,\downarrow}d_{m,\mu,\downarrow}^{}\rangle$,
i.e., the average $z$-th spin component vanishes. Then we inspect the
real parts of a subclass of complex order parameters
(which we get on convergence from the HF output), namely
$\langle d^\dagger_{m,\mu,\uparrow}d_{m,\mu,\downarrow}^{}\rangle$.
When the summation over $\mu$ is performed, i.e., if we calculate
$Re\big\{\sum_\mu\langle
d^\dagger_{m,\mu,\uparrow}d_{m,\mu,\downarrow}^{}\rangle\big\}$,
we obtain the value 0.52 which is just the average spin component along
the $x$ direction. The imaginary part of the same sum (here it is zero)
corresponds to the average spin component along $y$ direction.
This ends our digression.

\subsection{Zero-temperature ground state in YVO$_3$ }

The symmetry of YVO$_3$ at zero temperature is orthorhombic
\cite{Bla01,Ren03}. This corresponds to $C$-AO order accompanied by
$G$-AF$_z$ spin order. The bond lengths and average magnetization
values were reported in Refs. \cite{Kaw94,Nak99,Bla01,Ulr03,Ren03};
band gaps are $1.2-1.6$ eV \cite{Tsv04,Kum17}.

Our HF results on occupation numbers are virtually the same
(two digits accuracy) like those for LaVO$_3$ (shown in Table II).
As about spin-order just like it was shown in detail for LaVO$_3$ the
$z$ and $x$ easy spin directions are degenerate within 1 meV accuracy
(at least for our simplified geometries shown in Figs. 2 and 3).
The $T=0$ ground state for YVO$_3$ has $C$-AO order coexisting with
$G$-AF$_c$ spin-order and is best reproduced by HF results for
self-doping $x\approx 0.5$. For $x=1.0$ we find that the orbital order
vanishes.

\subsection{Zero-temperature ground state in BaVO$_3$}

To test how accurately the $d-p$ model works in the vanadium
perovskites we decided to test one more completely different case:
perovskite quasi-cubic BaVO$_3$ (with V$-$O bonds equal approximately
2.0 \AA), which is known to be a conductor \cite{Nis14} down to $T=0$.
This time we cannot use crystal-field splittings as the substance is
indeed very close to cubic, all octahedra are undistorted and therefore
$t_{2g}$ levels remain unsplit. The other significant difference
(with respect to LaVO$_3$) is that Ba cation donates not 3 but 2
electrons into one VO$_3$ unit.

With this input we run our computations only to find that for any
doping (including ideal-ionic picture with zero self-doping) and for
any starting conditions the obtained HOMO-LUMO gaps (after correcting
for virtual level shift) are negative. This is a clear indication that
BaVO$_3$ is a conductor in nice agreement with the experimental
findings. The same conclusion would be also reached for CaVO$_3$ ---
though CaVO$_3$ is not quasi-cubic and local crystal fields do split
$t_{2g}$ levels. Here the decisive factor is probably not symmetry but
the number of electrons transferred from a Ca cation to VO$_3$ unit
which is at most 2 (ideal ionic model) or (very likely) much
smaller, say within the $(1.0,1.5)$ interval.

\subsection{Remarks on high temperature ($T>77$ K) ground state of YVO$_3$}

First we should clearly state that for $T>0$ K the HF computations of
the ground state should not apply directly as we do not know the value
of entropy and do not determine the minimum of thermodynamic potential.
However just out of curiosity we did them anyway.

The bond lengths and average magnetization values were reported in
Refs. \cite{Kaw94,Nak99,Bla01,Ulr03,Ren03}; band gaps are $1.2-1.6$ eV
\cite{Tsv04,Kum17}. The symmetry of YVO$_3$ for $T>77$ K is monoclinic
\cite{Bla01,Ren03}.
Our HF occupation numbers we obtained  are very close to those shown in
Table II. The symmetry of the obtained ground state is $G$-AO order
with $C$-AF$_z$ spin order in accordance with the experiment.

The only disagreement with the experiment is that experimentally
\cite{Ulr03} the easy axis of magnetization is neither in $c$ direction
nor it is strictly located in the $ab$ plane; one finds spin components
of both types. Such a possibility was not investigated during our
computations. However, to be on defensive side, let us remind once
more that (like it was shown in detail for LaVO$_3$) the $z$ and $x$
easy spin directions are degenerate within 1.0 meV accuracy
(at least for our simplified geometries presented in Figs. 2 and 3).

\section{Summary and conclusions}
\label{sec:summa}

On some examples we have shown that the $d-p$ model is capable of
reproducing spin-orbital order in the vanadium perovskites.
The three basic fundamentals leading to non-zero orbital-order are:
(i) the electronic configuration of V ions which is close to $V^{3+}$;
(ii) non-zero local crystal fields (originating from collective JT
deformations) which split $yz/zx$ orbitals;
(iii) zero or small self-doping due to cations (i.e., electron donors
to the VO$_3$ lattice).
With these ingredients orbital order is generic --- it comes out
correctly for any reasonable Hamiltonian parameter set.

However the question \textit{what kind} of magnetic order accompanies
orbital order is more subtle. In particular different spin easy-axis
orientations are difficult to find as the states stable in HF
(candidates for being true ground state) are almost energetically
degenerate. In addition to this problem the stability and the type of
dominating magnetic order depends strongly on tiny effects occurring
on oxygens: small ($\pm 0.01$) spin modulations and small charge
modulations, i.e., $\pm(0.01-0.03)$. If one imposes same additional
assumptions (for example the assumption that oxygens in $ab$ planes are
nonmagnetic --- which may seem to be obvious but which is incorrect) in
hope that HF convergence will improve then the order in which the types
of magnetic order appear may even come out completely wrong.

The above problem (i.e., how to include tiny magnetization modulations
on oxygens) is non-existent for \textit{ab-initio} LDA or LDA with
local Coulomb interaction $U$ (LDA+$U$) approaches but at a cost of
many-fold increase of computational time and effort. On the other hand,
the $d-p$ model is not \textit{ab-initio} and HF computations performed
on the $d-p$ model can not reach the level of physical reliability such
as the LDA+$U$ does but still for extremely cheap and quick preliminary
computations in new perovskite materials with orbital and spin degrees
of freedom they are indeed of invaluable help.

Summarizing, the multi-band model considered here reproduces the
experimentally observed coexisting $G$-AO and $C$-AF spin order in the
vanadium perovskites. We emphasize that the minimal multi-band model
for the vanadium perovskites has to include all five $3d$ orbitals on
vanadium ions. Electron densities in $e_g$ orbitals are typically even
larger than that in the nominally empty third $t_{2g}$ orbital. This
redistribution of electron charge follows from rather strong $d-p$
hybridization with two $e_g$ orbitals which contribute to the total
electronic charge and magnetization of vanadium ions. Our calculations
suggest finite but rather low self-doping of $x=0.5$ in the vanadium
perovskites.

\acknowledgments
We kindly acknowledge support by Narodowe Centrum Nauki
(NCN, National Science Center) under Project
No.~2016/23/B/ST3/00839.

\end{document}